\documentclass[prl,showpacs,superscriptaddress,amsfonts,twocolumn]{revtex4}
\usepackage{bm}
\usepackage{amsfonts}
\usepackage{times}
\usepackage[dvips]{graphicx}
\usepackage[T1]{fontenc}
\usepackage{color}

\begin{document}

\title{Bailout Embeddings and Neutrally Buoyant Particles in Three-Dimensional 
Flows}

\author{Julyan H. E. Cartwright} 
\homepage{http://lec.ugr.es/~julyan}
\email{julyan@lec.ugr.es}
\affiliation{Laboratorio de Estudios Cristalogr\'aficos, CSIC, 
E-18071 Granada, Spain}
\author{Marcelo O. Magnasco} 
\homepage{http://asterion.rockefeller.edu/marcelo/}
\email{marcelo@sur.rockefeller.edu}
\affiliation{Mathematical Physics Lab, Rockefeller University, 
Box 212, 1230 York Avenue, NY 10021}
\author{Oreste Piro}
\homepage{http://www.imedea.uib.es/~piro}
\email{piro@imedea.uib.es}
\author{Idan Tuval.} 
\homepage{http://www.imedea.uib.es/~idan} 
\email{idan@imedea.uib.es} 
\affiliation{Institut Mediterrani d'Estudis Avan\c{c}ats, CSIC--UIB,
E-07071 Palma de Mallorca, Spain}

\date{Published as Phys. Rev. Lett. {\bf 89}, 264501, 2002.}

\begin{abstract}
We use the bailout embeddings of three-dimensional volume-preserving maps to
study qualitatively the dynamics of small spherical neutrally buoyant
impurities suspended in a time-periodic incompressible fluid flow. The 
accumulation of impurities in tubular vortical structures, the detachment of 
particles from fluid trajectories near hyperbolic invariant lines, and the 
formation of nontrivial three-dimensional structures in the distribution of 
particles are predicted.
\end{abstract}


\maketitle

The dynamics of small spherical particles immersed in a fluid flow have
received considerable attention in the past few years from both the theoretical
and experimental points of view. On one hand, these particles are the simplest
models for impurities whose transport in flows is of practical interest, and, 
on the other, their motion is governed by dynamical systems that even in the
most minimal approximations display a rich and complex variety of behavior.
When the density of the particles does not match that of the fluid, it is
intuitively clear that the trajectories of a particle and of a fluid parcel
will in general differ. This has been demonstrated in two-dimensional flows in
which particles with density higher than the basic flow tend to migrate away
from the parts of the flow dominated by rotation --- in the case of chaotic
flows, the KAM (Kolmogorov--Arnold--Moser) islands --- while particles lighter
than the fluid display the opposite tendency \cite{crisanti,tanga,partproc}. A
more surprising result, however, is that neutrally buoyant particles may also
detach from the fluid-parcel trajectories in the regions in which the flow is
dominated by strain, to settle in the KAM islands \cite{neutpartprl}. The
subtle dynamical mechanism responsable for the latter phenomenon has suggested
a method to target KAM islands in Hamiltonian flows, and a recent
generalization named a bailout embedding permits its extension to Hamiltonian
maps as well \cite{bailout1}. 

Despite its obvious importance from the point of view of applications, the case
in which the base flow is three-dimensional has been much less investigated.
The probable reason for this is that very few simple realistic
three-dimensional incompressible flow models exist. The few that are simple are
not realistic --- e.g., the ABC (Arnold--Beltrami--Childress) flow
\cite{dombre} --- and those that are realistic are far more complex. In the
study of the Lagrangian structure of three-dimensional incompressible
time-periodic flows, where this difficulty is already present, the alternative
approach of qualitatively modeling the flows by iterated three-dimensional
volume-preserving (Liouvillian) maps has been successful in predicting
fundamental structures later found both in more realistic theoretical flows and
in experiments \cite{feingold88.2,pirofein,spheresletter,spherespaper}.
However, no similar approach has been followed to describe the motion of
impurities in this kind of flow.

This Letter introduces the idea of bailout embeddings to non-Hamiltonian 
systems. Two fundamentally unexplored problems are attacked with this new 
technique: firstly the investigation of generic structures of the dynamics of 
small particles in three-dimensional time-periodic flows, and secondly the 
study of the effects of noise and fluctuations on the dynamics of such 
particles, as well as the practical use of this analysis to reveal structures 
in three-dimensional flows that would be largely hidden to other methods.
Regarding the former, we produce qualitative predictions using the
bailout approach for maps that we confirm for flows. For the
latter, we show that intricate but well defined structures may arise in a
surprising way in the distribution of particles driven by extremely chaotic
flows and the presence of noise. Bailout embedding is amenable to theoretical
and experimental analysis in fluid dynamics, but the technique has many
potential uses in the much broader community of physicists dealing in one way
or another with nonlinear dynamical systems.

\begin{figure*}[ht]
\begin{center}
\includegraphics*[width=0.9\textwidth]{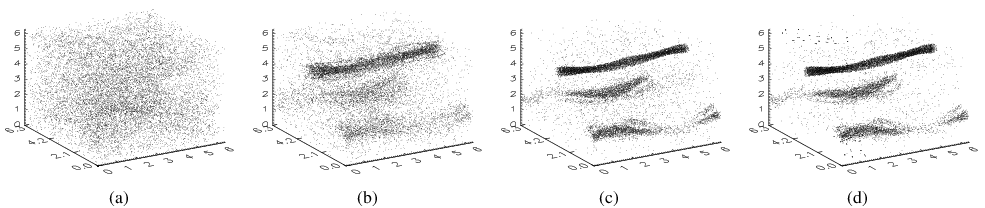}
\end{center}
\caption{\label{baildraw} 
For a homogeneous distribution of one hundred initial conditions
we plot only the last 1000 steps of the map evolution for different values of
$\lambda$: (a) $\lambda=2$, (b) $\lambda=1$, (c) $\lambda=0.6$, 
(d) $\lambda=0.5$. The images represent the $[0,2\pi]$ cube in the phase
space.}
\end{figure*}

Let us first recall the equation of motion for a small, neutrally buoyant,
spherical tracer in an incompressible fluid (the Maxey--Riley equation)
\cite{maxey,michaelides}. Under assumptions allowing us to retain just the
Bernoulli, Stokes drag, and Taylor added mass contributions to the force
exerted by the fluid on the sphere (the particle 
radius and its Reynolds number are small, as are the velocity gradients
around it \cite{neutpartprl}), the equation of motion for the particle at
position $x$ is
\begin{equation}  
\frac{d}{dt}(\dot{\bm x}-{\bm u}({\bm x}))=-(\lambda+{\bm\nabla\bm u})
\cdot(\dot{\bm x}-{\bm u}({\bm x}))
,\label{part_in_fluid}
\end{equation}
where $\dot{\bm x}$ represents the velocity of the particle, $\bm u$ that of
the fluid, $\lambda$ a number inversely proportional to the Stokes number of
the particle --- the ratio of the relaxation time of the particle back onto the
fluid trajectories to the time scale of the flow --- and
$\bm\nabla\bm u$ the Jacobian derivative matrix of the flow.
The difference between the particle velocity and the velocity of the
surrounding fluid is exponentially damped with negative damping term
$-(\lambda+\bm\nabla\bm u)$. In the case in which the flow gradients reach the
magnitude of the viscous drag coefficient, there is the possibility that
the Jacobian matrix $\bm\nabla\bm u$ may acquire an eigenvalue of positive real
part in excess of the drag coefficient. If in these instances we can discard 
the time dependence of the eigenvectors it is clear that the trajectories, 
instead of converging exponentially onto those defined by
$\dot{\bm x}=\bm u$, may detach from them.

For incompressible two-dimensional
flows, since the Jacobian matrix is traceless, the two eigenvalues must add up
to zero, which implies that they are either both purely imaginary or both purely
real, equal in absolute value and opposite in sign. The result is that
the particles can abandon the fluid trajectories in the neighborhood of the
saddle points and other unstable orbits, where the Jacobian eigenvalues are
real, and eventually overcome the Stokes drag, to finally end up in a regular
region of the flow on a KAM torus dominated by the imaginary eigenvalues. From
a more physical point of view, this effect implies that the particles tend to
stay away from the regions of strongest strain.
In contrast to the two-dimensional case, in time-dependent three-dimensional
flows the incompressibility condition only implies that the sum of the three
independent eigenvalues must be zero. This less restrictive condition allows
for many more combinations. Triplets of real eigenvalues, two positive and one
negative or vice versa, as well as one real eigenvalue of either sign together
with a complex-conjugate pair whose real part is of the opposite sign, are
possible. Accordingly, chaotic trajectories may have one or two positive
Lyapunov numbers, and a richer range of dynamical situations may be expected.

Instead of investigating all these in terms of a given fully fledged
three-dimensional time-periodic model flow, we follow a qualitative
approach based on iterated maps that roughly reproduces the properties of the
impurity dynamics in a generic flow of this kind. In order to construct the map
we first note that the dynamical system governing the behavior of neutrally
buoyant particles is composed of some dynamics within another larger set of
dynamics. Equation~(\ref{part_in_fluid}) can be seen as an equation for a
variable $\bm\delta=\dot{\bm x}-\bm u$ which in turn will define the
equation of motion $\dot{\bm x}=\bm u$ of a fluid element whenever the
solution of the former be zero. In this sense we may say that the fluid parcel
dynamics is embedded in the particle dynamics. In reference to the fact that
some of the embedding trajectories abandon some of those of the embedded
dynamics, the generalization of this process is dubbed a bailout embedding
\cite{bailout1}.

It is rather easy to construct this type of embedding for map dynamics. 
Given a map $\bm X_{n+1}=\bm T(\bm X_{n})$,
a general bailout embedding is given by
\begin{equation}\label{bailout_map}
\bm X_{n+2}-\bm T(\bm X_{n+1})=\bm K(\bm X_{n})
\cdot(\bm X_{n+1}-\bm T(\bm X_{n}))
,\end{equation}
where $\bm K(\bm X)$ is the bailout function whose properties
determine which trajectories of the embedded map will be eventually
abandoned by the embedding.
The particular choice --- naturally imposed by the particle dynamics ---
of the gradient as the bailout function in a flow
translates in the map setting to
\begin{equation}\label{bail_function}
\bm K(x)=e^{-\lambda }\cdot\bm \nabla \bm T
.\end{equation}
Bailout embeddings have been used to investigate targeting of KAM tori in
Hamiltonian systems, as well as to explore generic properties of the
distribution of small particles immersed in incompressible two-dimensional
fluid flows \cite{bailout1}, which are also of a Hamiltonian nature.

Here we consider the bailout embedding of a
class of non-Hamiltonian systems: three-dimensional volume-preserving maps.
In particular, we choose to represent qualitatively chaotic three-dimensional
incompressible base flows that are periodic in time by ABC maps, a family
\begin{equation}
\label{Tabc}
\bm T=\bm T_{ABC}: (x_n,y_n,z_n)\longrightarrow(x_{n+1},y_{n+1},z_{n+1})
,\end{equation}
where
\begin{eqnarray}
\label{Tabc_explicita}
x_{n+1} &=&x_n+A\sin z_n+C\cos y_n\,(\bmod 2\pi), 
\nonumber \\ 
y_{n+1} &=&y_n+B\sin x_{n+1}+A\cos z_n\,(\bmod 2\pi),
\\ 
z_{n+1} &=&z_n+C\sin y_{n+1}+B\cos x_{n+1}\,(\bmod 2\pi),
\nonumber 
\end{eqnarray}
that displays all the basic features of interest of the evolution of fluid
flows. Depending on the parameter values, this map possesses two
quasi-integrable behaviors: the one-action type, in which a KAM-type theorem
exists, and with it invariant surfaces shaped as tubes or sheets; and the
two-action type displaying the phenomenon of resonance-induced diffusion
leading to global transport throughout phase space \cite{feingold88.2}.

Let us now study the dynamics defined by 
Eqs~(\ref{bailout_map})--(\ref{Tabc_explicita}). 
We first concentrate on the cases in
which the flow is dominated by one-action behavior. In these we find an
interesting generalization of the behavior already seen in two dimensions:
particles are expelled from the chaotic regions to finally settle
in the regular KAM tubes. As an example, we take values of $A$, $B$, and $C$
that lead to almost ergodic behavior of the fluid map: a single fluid
trajectory almost completely covers the phase space. However, from randomly
distributed initial conditions, the particle trajectories inevitably visit some
hyperbolic regions where they detach from the corresponding fluid trajectory.
In this fashion they find their way inside the invariant elliptic structures
where they can finally relax back onto a safe fluid trajectory. In
Fig.~\ref{baildraw} we show how a homogeneous distribution of 
particles in the fluid flow, after a large number of stabilization iterations,
finally settles inside the tubular KAM structures for different values of the
parameter $\lambda$. When the value of $\lambda$ decreases, more random
trajectories follow this evolution: more particles fall into the
invariant tubes.

\begin{figure}[tb]
\begin{center}
\includegraphics*[width=0.8\columnwidth]{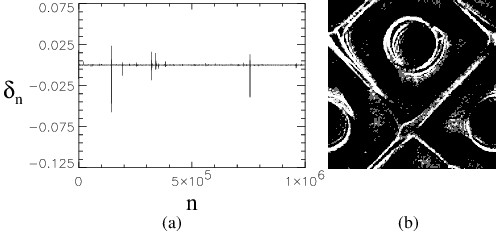}
\end{center}
\caption{\label{2acciones}
(a) The evolution of one component of the difference $\bm\delta_n$ between
the particle and the fluid-flow velocities as a function of the number of
iterations. (b) An $xy$ slice of the phase space with those points where 
$\bm\delta_n$ is greater than $\bm\delta_0$. The square is the region 
$[0,2\pi]\times[0,2\pi]$.}
\end{figure}

In the two-action case, the eigenvalues of the Jacobian are very small on large
portions of the trajectory, so that separation may only occur sporadically
during the short time intervals in which the fluid parcel crosses the 
fast-motion resonances \cite{feingold88.2}. Most of the time, particles and
fluid parcels follow exponentially convergent trajectories, causing the
separations to be practically unobservable except for very small values of
$\lambda$. Most probably, once the particles converge to the fluid dynamics,
they remain attached. However, by adding a small amount of white noise, we can
continually force the impurity to fluctuate around the flow trajectory
\cite{bailout2}. From the application point of view, this noise may be
considered to represent the effect of small scale turbulance, thermal
fluctuations, etc., but here we use it only as a dynamical device. With
this, the particles arrive in the neighborhood of the resonances with a
non-negligible velocity mismatch with the fluid that is considerably amplified
during the transit across the resonance. The measure of this mismatch is then a
good detector of the proximity of the resonance.

Consider the following stochastic iterative system: 
\begin{equation}
\label{bail_map}
\bm X_{n+2}-\bm T(\bm X_{n+1})=e^{-\lambda}\bm\nabla\bm T
\cdot(\bm X_{n+1}-\bm T(\bm X_n))+\bm\xi_n 
,\end{equation}
where the noise term $\bm\xi_n$ satisfies $\langle\bm\xi_n\rangle=0$, and 
$\langle\bm\xi_n\,\bm\xi_m\rangle =
\varepsilon(1-e^{-2\lambda})\,\delta_{mn}\,{\mathrm I}$.
We can recast Eq.~(\ref{bail_map}) into
\begin{eqnarray} 
\label{bail_map2}
\bm x_{n+1}&=&\bm T(\bm x_n)+\bm \delta_n, \nonumber \\
\bm\delta_{n+1}&=&e^{-\lambda}\bm\nabla \bm T\cdot\bm\delta_n+\bm\xi_n,
\end{eqnarray}
if we define the velocity separation between the fluid and the particle
as $\bm\delta_n=\bm x_{n+1}-\bm T(\bm x_n)$.
We illustrate the behavior referred above by studying the most ergodic
two-action case, in which all the fluid trajectories intersect the resonant
lines. In Fig.~\ref{2acciones}a we show how $\bm\delta_n$ grows strongly at
some points, which correspond to the crossings of the resonant lines. In
Fig.~\ref{2acciones}b we plot an $xy$ slice of the
three-dimensional cube, choosing those points where the value of 
$\bm\delta_n$ is greater than a minimum value $\bm\delta_0$. As shown,
we recover the resonant structure previously noted \cite{feingold88.2,pirofein}.

\begin{figure}[tb]
\begin{center}
\includegraphics*[width=0.8\columnwidth]{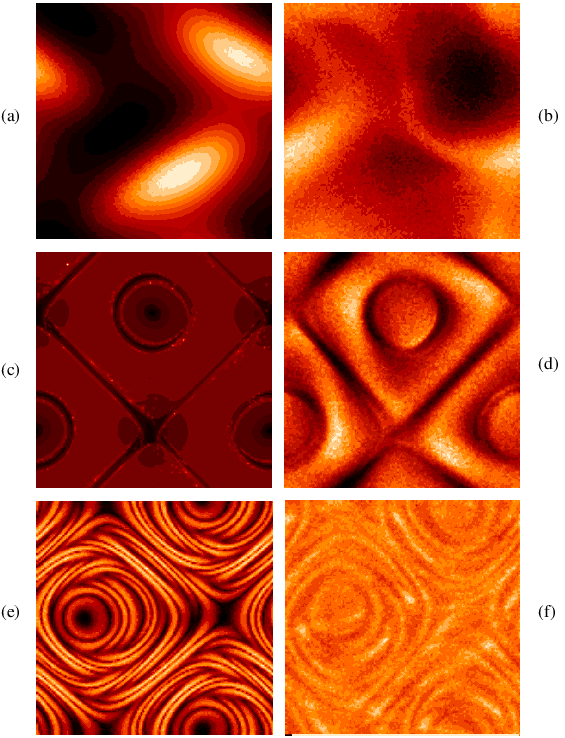}
\end{center} 
\caption{\label{temp}
The temperature amplitude --- lighter is hotter ---
for the one-action, two-action and most chaotic cases
((a), (c), and (e) respectively), together with the corresponding slices of the
impurity dynamics (histogram) in the phase space ((b), (d), and (f)). All
images are the $[0,2\pi]\times[0,2\pi]$ region in the $xy$ axis, for a
slice in the $z$ direction corresponding to the values $z\in[0,0.49]$.} 
\end{figure}

This is the most primitive way to obtain useful information from the noisy
particle dynamics. A shrewder analysis \cite{bailout2}
shows how the variance of the separation $\bm\delta_n$ between particles and
fluid trajectories and the variance of the noise $\bm\xi_n$ are related by a
function that depends only on the particular point of the phase space that we
look at, in a sort of temperature amplitude for the fluctuations of $\bm\delta$,
\begin{equation}
\label{temperatura}
\mathcal{T}(x)=\frac{\left\langle\bm\delta^2\right\rangle}
{\left\langle\bm\xi^2\right\rangle}=\sum_{j=0}^\infty 
\left( e^{-j\lambda}\cdot 
\prod_{k=0}^j\bm\nabla\bm T\left|_{T^{-k}(x)}\right.\right)^2
.\end{equation}
This amplitude takes different values at different points of the flow. At those
points that the particle dynamics tries to avoid, its value increases, so
the particle prefers to escape the hot regions and to fall into the
cold ones. 

In Fig.~\ref{temp} we illustrate this phenomenon. First we analyze a one-action
situation showing the temperature amplitude, as well as the impurity dynamics;
Figs~\ref{temp}a and \ref{temp}b respectively. Again we use slices of the
three-dimensional cube to show the situation more clearly. Fig.~\ref{temp}a
shows the temperature in a scaled color code. 
Fig.~\ref{temp}b shows a
histogram of visits that a single particle pays to each bin of the space. The
agreement between the higher-temperature regions and the less-visited ones is
evident. Next we plot the same pictures but in the two-action case studied
before; Figs~\ref{temp}c and \ref{temp}d. Finally, we apply this analysis to a
generic chaotic case where we do not have any information about the phase space
structure. We show in Fig.~\ref{temp}e how the invariant manifolds are very
twisted, and in Fig.~\ref{temp}f how the particles, even so, try to find the
coldest regions of the flow.

In order to confirm that the above-described behavior is not an artefact of our
mapping-based approach, we have performed analogous simulations using a
continuous-time model as a base flow. We have considered neutrally
buoyant particles immersed in a modified version of the ABC flow, 
\begin{eqnarray} \label{ABC-time-dependent}
\dot{x}&=&(1+\sin\ 2\pi t)\cdot(A\sin z+C\cos y), \nonumber \\
\dot{y}&=&(1+\sin\ 2\pi (t+1/3))\cdot(B\sin x+A\cos z), \\
\dot{z}&=&(1+\sin\ 2\pi (t+2/3))\cdot(C\sin y+B\cos x), \nonumber
\end{eqnarray}
in which each
component of the velocity vector field is sinusoidally modulated with a
relative phase shift of $2\pi/3$, and where $x$, $y$, and $z$ are to be 
considered modulo $2\pi$.
While a detailed analysis of the dynamical aspects of this flow is beyond
the scope of this Letter, we advance that it shows structures similar
to those of the ABC maps, i.e., a complex array of KAM sheets and tubes
surrounded by chaotic volumes. Neutrally buoyant particles evolved
according to the true (simplified) Maxey--Riley equations,
Eq.~(\ref{part_in_fluid}), based on this flow, show exactly
the same tendency to accumulate inside KAM tubes as in the map case.

This application of bailout embedding is the first to be reported for a
non-Hamiltonian dynamical system. Our approach can be pursued with two
different goals in mind: on one hand, it contributes to the understanding of
the physical behavior of impurities, and on the other, it provides a
mathematical device to learn about the dynamical structures of the base flow in
situations where these are very difficult to elucidate directly. Both bodies of
information are important to improve our presently scant knowledge of the
transport properties of three-dimensional fluid flows.

This research was performed by IT as part of his doctoral thesis under the
supervision of OP, together with additional input from JHEC and MOM. JHEC
acknowledges the financial support of the Spanish CSIC, Plan Nacional del
Espacio contract PNE-007/2000-C, MOM acknowledges the support of the Meyer
Foundation, and OP and IT acknowledge the Spanish Ministerio de Ciencia y
Tecnologia, Proyecto CONOCE, contract BFM2000-1108 and Proyecto IMAGEN,
contract REN2001-0802-C02-01.

\bibliography{database}

\end{document}